# Effects of point defects on oxidation of 3C-SiC

Jianqi Xi[‡,*], Cheng Liu[†,*], Izabela Szlufarska[‡†§]

[‡] University of Wisconsin-Madison, Department of Material Science and Engineering, 1509 University Ave., Madison, WI, 53706, U.S.A.
[†] University of Wisconsin-Madison, Department of Engineering Physics, 1500 Engineering Dr., Madison, WI, 53706, U.S.A.

* Co-first author.
§ Correspondence should be addressed to: szlufarska@wisc.edu

The influence of implantation-induced point defects (PDs) on SiC oxidation is investigated via molecular dynamics simulations. PDs generally increase the oxidation rate of crystalline grains. Particularly, accelerations caused by Si antisites and vacancies are comparable, and followed by Si interstitials, which are higher than those by C antisites and C interstitials. However, in the grain boundary (GB) region, defect contribution to oxidation is more complex, with C antisites decelerating oxidation. The underlying reason is the formation of a C-rich region along the oxygen diffusion pathway that blocks the access of O to Si and thus reduces the oxidation rate, as compared to the oxidation along a GB without defects.

## 1. Introduction

Due to its excellent high-temperature strength, superior radiation tolerance, and good chemical stability, silicon carbide (SiC) has been proposed as an attractive candidate for a variety of applications. For example, it has been used in high-power and high frequencies devices, such as metal-oxide-semiconductor field-effect transistors [1]. In addition, SiC and SiC/SiC composites have long been known for their potential as structural components in fusion and fission reactors [2-4]. In these applications, doping via ion-implantation and neutron irradiations inevitably induce lattice defects. These defects not only deteriorate physical and mechanical properties of SiC, such as the thermal conductivity and strength [3, 4], but they may also change the chemical stability, leading to a suppressed corrosion resistance.

Historically, radiation-enhanced corrosion has been investigated primarily in metals, and it has been attributed to radiation-enhanced diffusion and segregation [5]. For instance, Was *et al.* reported that the radiation-induced Cr depletion is responsible for the observation of intergranular stress corrosion cracking in irradiated stainless steels [5]. In the case of SiC, it had been previously suggested that radiation-enhanced corrosion would be limited [6], likely because of the relatively good radiation and corrosion resistance of this material. However, recent experiments on SiC demonstrated that radiation can accelerate corrosion of SiC in hydrothermal water [7, 8] or in molten salts [9, 10]. For example, Kondo *et al.* studied the influence of radiation-induced defects on the hydrothermal oxidation/corrosion of SiC in autoclave testing with high content of oxygen [7, 8]. They found that dissolution rate of oxides formed on an irradiated SiC surfaces is orders of magnitude higher than that on an unirradiated SiC, which was hypothesized to be due to enhanced oxygen diffusion within the irradiated samples. Similar results have been observed for corrosion in molten salts [10]. In their study, Li *et al*. found that after placing the irradiated/unirradiated SiC samples in molten salt for 160 h, the irradiated sample was more heavily corroded than the unirradiated sample [10]. These results indicated that radiation-induced defects could play a significant role in surface oxidation/corrosion of SiC.

In addition, recent experimental and theoretical investigations have found that grain boundaries (GBs) in SiC are more vulnerable to oxidation [11, 12]. By exposing an unirradiated SiC to the supercritical water, Tan *et al*. have found that incoherent GBs are more significantly corroded than the coherent ones [11]. These experimental results are consistent with our recent molecular dynamics (MD) simulations [12]. Specifically, we have previously shown that incoherent tilt GBs oxidize faster than the bulk SiC whereas coherent GBs do not. This result was explained by the distribution of the local strain and of the under-coordinated Si within the GB region, both of which reduce the positive charge on Si atoms, rendering them more reactive with oxygen [12]. These earlier studies suggested that oxidation at SiC GBs depends strongly on the local atomic structure.

Local atomic structure of SiC can be altered by irradiation. Particularly, the GB regions in materials are often regarded as sinks for radiation-induced defects, and it has been recently shown that radiation-induced segregation to GB regions in SiC can lead either C enrichment or C depletion at GBs, depending on temperature [13]. Changes in the GB microchemistry could further affect mechanisms and rates of oxidation along GBs. Here, we carry out classical MD simulations of defective SiC in contact with oxygen in order to determine how irradiation-induced defects affect oxidation, especially at GBs. More specifically, we use reactive force field ReaxFF developed by van Duin *et al.* [14] for oxidation of SiC to determine the influence of point defects on both intragranular and intergranular oxidation of SiC.

## 2. Methodology

Classical MD simulations have been performed to study the influence of defects on intragranular and intergranular oxidation of 3C-SiC by using single crystal and bi-crystal configurations. For single crystals, our previous MD simulations have shown that the oxidation rate of SiC has slight dependence on surface orientation but the difference between oxidation rates of different surfaces is small relative to the difference between oxidation rates of any single crystal surface and the incoherent GB. To reduce the computational cost, here we focus only on the surface with the (100) crystallographic orientation but the qualitative trends with defect types and defect concentrations are not expected to depend on the crystal orientation. Considering the dominance of incoherent GBs in engineered 3C-SiC and the fact that they provide fast oxidation pathways [12], in this study we perform simulations on bi-crystals with incoherent GBs. Models of incoherent GBs were built by combining two single crystals with different crystallographic orientations along the *z* directions (see Fig. 1). At the same time, depending on the surface termination along *z* directions, the interface at the GBs can be either Si-rich or C-rich. Here, to save the computational cost, we only considered the incoherent GB formed by combining crystals with (100) and (111) orientations and Si-termination along *z* directions. For each sample, the dimensions along the *x* and *z* directions are 30 and 50 Å, respectively, whereas dimensions along the *y* direction are 60 and 120 Å for single crystal and bi-crystal, respectively.

To determine the role of point defects in SiC oxidation, we considered six different types of point defects: silicon and carbon interstitials ($I_{Si}$ and $I_C$), silicon and carbon vacancies ($V_{Si}$ and $V_C$), silicon antisite ($Si_C$) with silicon occupying carbon sublattice, and carbon antisite ($C_{Si}$) with carbon occupying silicon sublattice. In this work, configurations of the most stable interstitials in our MD simulations are consistent with those predicted in density functional theory calculations [15]. Namely, the dumbbell carbon interstitial ($I_C$), where two C atoms share a regular carbon sublattice along <100>, i.e., C-C<100>, and silicon interstitial ($I_{Si}$), which is located at the tetrahedral position surrounded by four regular C atoms, i.e., $Si_{TC}$. All these point defects were uniformly and randomly distributed within each single crystal sample and within 10 Å thick regions centered at each incoherent GB. After constructing such models, we equilibrated the samples at 300 K for 2.5 ps. Subsequently, the samples were relaxed at 500 K for 2.5 ps, and then quenched to 300 K over 5 ps in isothermal-isobaric (NPT) ensemble at zero pressure.

Once the samples were prepared, oxygen molecules with the number density of 0.02/Å³ were randomly and uniformly inserted into a 40 Å vacuum region above the SiC surface; the corresponding gas pressure at 2000 K is ~552.5 MPa. Our simulated oxygen partial pressure is significantly higher than pressures used in dry thermal oxidation experiments, which is in the order of tens of kPa [12]. This high pressure used in our simulations is necessary to observe reactions on the time scales of MD simulations. Similar pressures were used in earlier MD simulations of SiC oxidation, where it was shown that such simulations give a reasonable description of chemical reactions [14, 16, 17]. Periodic boundary conditions were applied along *x* and *y* dimensions of the simulation cell. A reflective boundary is used at the top of the simulation box, along *z* direction, in order to prevent oxygen from leaving the simulation cell. In order to maintain the position of the bi-crystal during the subsequent simulations, atoms within the bottom of 10 Å were not allowed to relax. The entire system was then equilibrated at 300 K for 2.5 ps, followed by heating to 1500 K within 2.5 ps, and finally annealed at 1500 K for 200 ps. Evolution of oxidation was recorded from the beginning of heating. Temperature was controlled using the Nose-Hoover thermostat [18]. The timestep of 0.25 fs and the temperature damping constant of 25 fs were used in the simulations. All simulations were performed using ReaxFF [14, 16], as implemented in Large-scale Atomic/Molecular Massively Parallel Simulator (LAMMPS) package [19]. The oxide thickness is defined as the difference between the average height of the top three silicon atoms and the average height of the deepest three oxygen atoms. We have tested that the choice of the number of atoms (different from three) does not change the qualitative trends reported in this paper. The thickness defined with this method represents the average oxide thickness, as discussed in Ref [12]. In the bicrystal, we report the average oxide thickness within a 10 Å thick region centered at the boundary.

## 3. Results
### 3.1 Effect of PDs on oxidation of single crystal SiC

The effect of point defects on simulated oxidation of crystalline SiC has been characterized by measuring the evolution of oxide thickness during the initial stage of oxidation (~200 ps), as shown in Fig. 2. We can see that all samples (with and without defects) show a logarithmic dependence of the oxide thickness on time during the very early stages of oxidation (~100 ps) [12]. During this period oxygen molecules are first adsorbed and dissociated onto the SiC surface, and then the dissociated oxygen atoms react with surface Si atoms to form a silica oxide layer. These results suggest that the existence of defects does not change the initial oxidation mechanisms in crystalline SiC. Nevertheless, from Fig. 2(a), we can clearly see that, regardless of the defect species, oxidation rate in defective samples is accelerated relative to undefected samples, indicating defect-enhanced oxidation of crystalline SiC.

Our simulations also reveal that not all point defects contribute equally to acceleration of oxidation. Such information is difficult to obtain from experiments where contributions from different defect types cannot be isolated. As shown in Fig. 2, for the same defect concentration, the effects of $Si_C$ and vacancies (either $V_C$ or $V_{Si}$) on oxidation of crystalline samples are comparable to each other and higher than the effects of other defects. Next, in terms of the magnitude of the effect, are Si interstitials, and the smallest effect (although not negligible) comes from C interstitial and C antisites. This finding can be understood when one considers the affinity of oxygen for silicon. The presence of Si antisite (Si atoms on C sublattice) locally increases the probability of Si-O interaction. Analysis of charge distribution shows that within the Si enriched region, the effective charges of Si atoms are significantly less positive (i.e., more electrons belong to these Si atoms) than those in perfect SiC (~0.574 |e|), see Fig. 3. Excess electrons localized around Si atoms would destabilize the Si-C covalent bonds and make these Si atoms more reactive with oxygen, and thus accelerate the oxidation process. Similar results can also be found in other defect cases, i.e., the effective charges of Si atoms in these defective systems are generally less positive than those in perfect SiC, as shown in Fig. 3. In addition to Si charge reduced (more negative) by the presence of a defect, the existence of vacancies creates broken bonds (undercoordinated atoms) as well as increases a local strain. The effects of reduced charge on Si and of strain cannot be simply separated from each other, i.e., applying strain to a perfect SiC crystal leads to redistribution of atomic charges, as demonstrated previously by MD simulations [12].

We have also found that the rate at which oxidation is accelerated with increasing defect concentration (see Fig. 2(b)). To demonstrate this trend, we plot the oxide thickness of samples with different concentrations of defects. The main reason for considering vacancies (of both types) and $Si_C$ antisites is that (as can be seen from Fig. 2(a)) these defects have the largest effect on oxidation. More broken bonds (introduced by vacancies) and excess Si atoms make the system more reactive. Based on the understanding of the physics controlling these trends, we expect that the effect of concentration will be similar for other defects.

**3.2 PDs effects on grain boundaries oxidation**

In order to test the influence of defects on the intergranular oxidation rate, we have studied oxidation of bi-crystal samples with point defects randomly inserted within a 10 Å wide region centered at each incoherent GBs, as shown in Fig. 4. Incoherent GBs have been previously shown to accelerate oxidation relative to single crystal due to the presence of strain and poorly-coordinated Si atoms in the GB region, even without irradiation-induced point defects [12]. In Fig. 4(a), we show how oxide grows in such incoherent GBs when they contain additional point defects, as would be introduced by radiation. Similarly as in the case of oxidation of single crystal SiC, the existence of Si antisites and Si interstitials, as well as vacancies accelerates oxidation at GBs. However, it is surprising to find that C antisites actually suppress the oxidation rate at the GB, which is the opposite effect to what we have seen in the single crystal (see Fig. 2). C interstitials also have a different effect at the GB than in a single crystal. At the GB, these defects initially accelerate oxidation (in the first ~150 ps), but then the oxidation rate becomes comparable to the samples without defects.

Why does the excess of C atoms at GB suppress oxidation rate or, at a minimum, prevents acceleration of oxidation that is expected for systems containing defects? All defects destabilize bonds through strain, and/or change in the coordination number of atoms, which both result in redistribution of charge (see Fig. 3). In a crystalline grain that means that all defects accelerate oxidation, as shown in Fig. 2. In GBs this trend is counter-balanced by the fact that oxygen has a stronger affinity to Si, and the excess C at the GB leads to an increased fraction of C-C bonds (the fractions of C-C bonds in the GB region with 10% of $C_{Si}$, $I_C$, $Si_C$, $I_{Si}$ are ~6.98%, ~3.67%, ~0.44%, and ~0.46%, respectively), which blocks the access of oxygen to Si atoms along the narrow GB path. The effect is stronger for C antisites (which remove Si) than for C interstitials. Depending on the specific C defect state and its concentration, the acceleration of oxidation expected for this type of defect relative to the pure incoherent GB is either suppressed or even decelerated. The key to radiation-accelerated oxidation is that defects introduced in the process do not interrupt a continuous path of Si atoms that can react with oxygen.

## 4. Discussion and conclusion

In this work, we performed MD simulations to determine the influence of radiation-induced point defects on intragranular and intergranular oxidation in SiC. Defect-accelerated oxidation, regardless of the type of defect, is generally observed within crystalline SiC. This effect is due to destabilization of bonds because of strain or poor coordination of atoms and it is reflected in a reduced charge on Si atoms. The magnitude of the effects is the largest for Si antisites and all vacancies, then for Si interstitials, and the slowest for C defects, such as C interstitials and C antisites. In contrast, when defects segregate to or directly form at the GB, they can either accelerate or suppress GB oxidation, depending on the defects type. Our results suggest that C enrichment of GBs (e.g., either during irradiation or by processing) would slow down GB oxidation. Interestingly, it has been recently reported that high-energy GBs in SiC grown by chemical vapor deposition (CVD) are depleted in carbon [13], which according to our results means that they would be more vulnerable to oxidation than stoichiometric GBs. At the same time radiation of SiC has been shown to lead to enrichment of GBs in C, which suggests that while radiation can accelerate oxidation in crystalline grains, it could actually reduce oxidation rates at GBs if C enrichment is significant.

## Acknowledgment

The authors gratefully acknowledge financial support from the US Department of Energy Basic Energy Science Grant # DE-FG02-08ER46493.

Fig. 1.

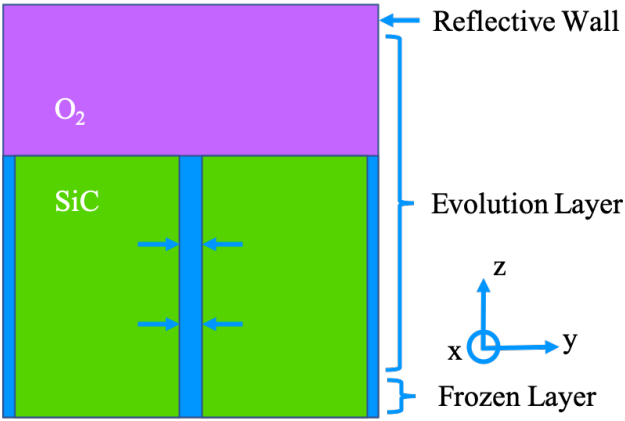

Fig. 2.

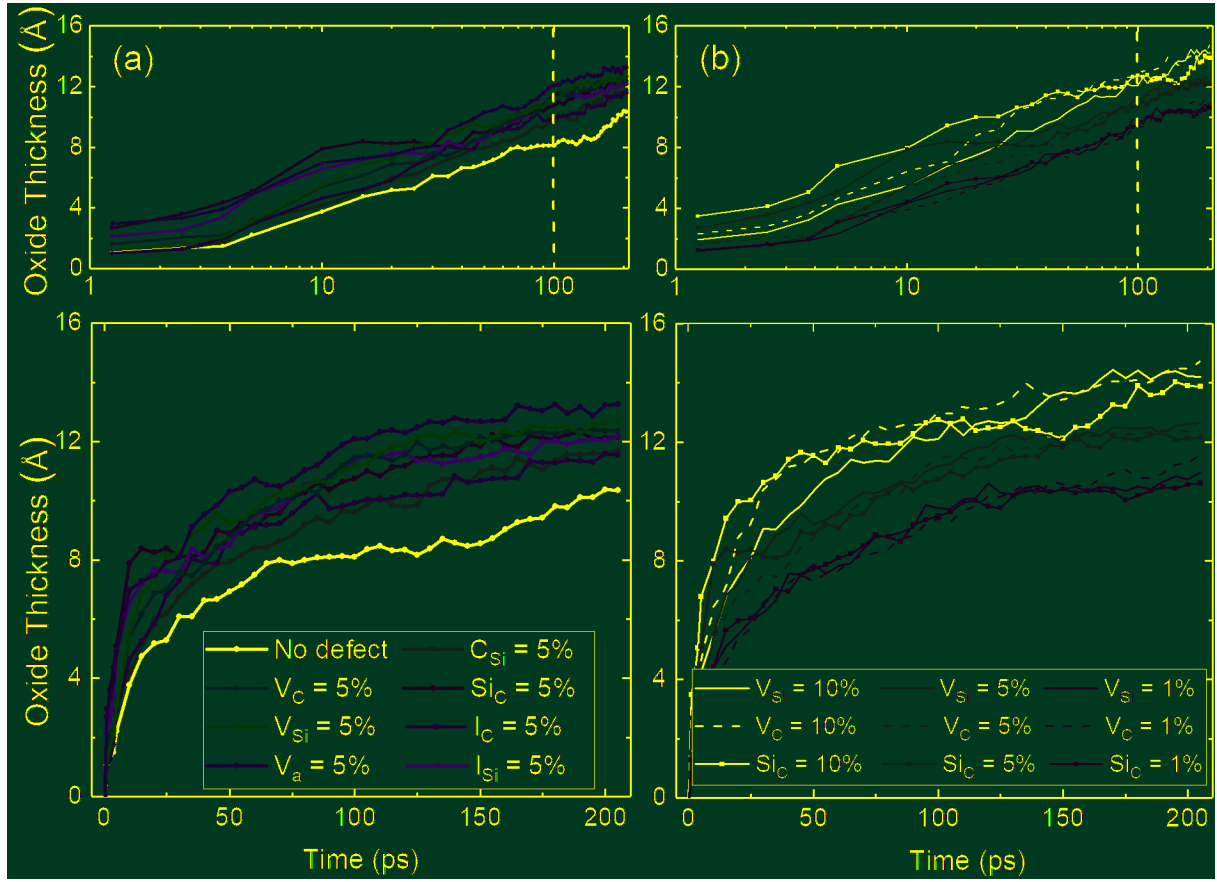

Fig. 3

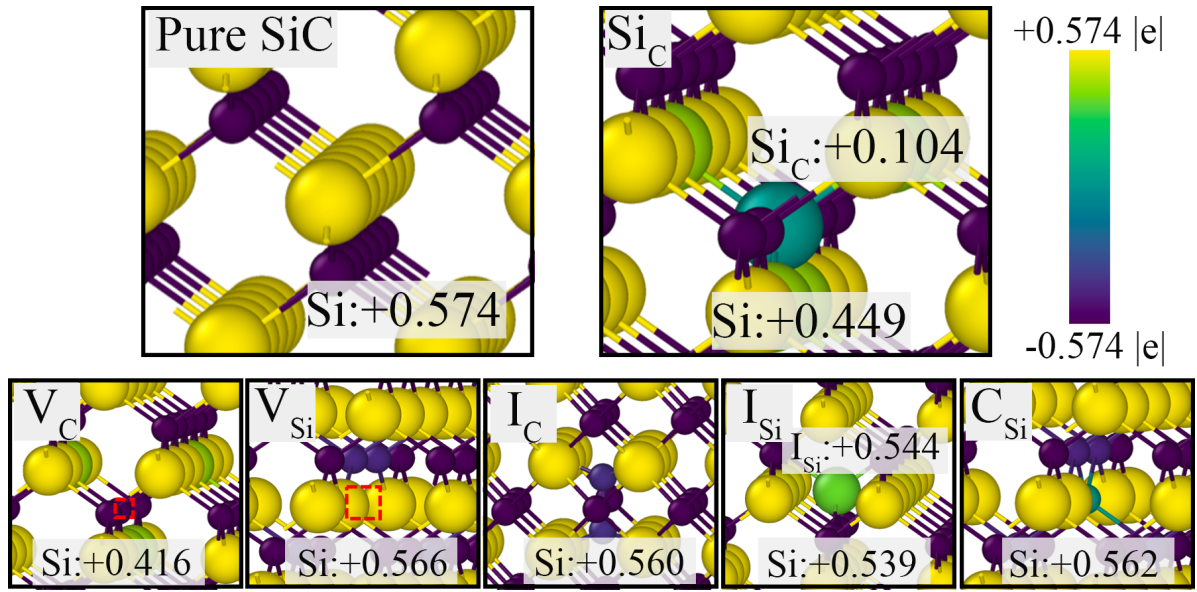

Fig. 4

Fig. 1. Schematic of a simulation cell with an incoherent GB. The purple region is the $O_2$ layer, and the green region is the SiC bulk. GB is inserted in the middle of the simulation cell which is represented by the blue region.

Fig. 2. (a) Evolution of oxide thickness on crystalline SiC without and with PDs at 5% concentrations. (b) Evolution of oxide thickness on crystalline SiC with different concentrations of vacancies and $Si_C$ antisites. $V_a$ denotes the vacancies mixed with the equal amount of C and Si vacancies in sample. E.g., $V_C$=5% indicates that 5% of C sites are empty, and $V_a$=5% means that 5% of C sites and 5% of Si sites are empty.

Fig. 3. Charge distribution of nearest-neighbor Si atoms around each point defect. The unit of effective charge is |e|.

Fig. 4. (a) Evolution of oxide thickness within incoherent GB region without and with PDs at 10% concentration. (b) Comparison of the effect of carbon antisite ($C_{Si}$) and silicon antisite ($Si_C$) on oxide thickness growth within incoherent GB region at 10% and 15% concentrations.